**Comparative prospects of imaging methods for whole-brain mammalian connectomics**

Logan Thrasher Collins,[1,*] Todd Huffman, Randal Koene[2]

[1]Washington University in St. Louis, Department of Biomedical Engineering
[2]Carboncopies Foundation
[*]Corresponding author, clogan@wustl.edu

**Abstract:** Mammalian whole-brain connectomes are a foundational ingredient for holistic understanding of brains. Indeed, imaging connectomes at sufficient resolution to densely reconstruct cellular morphology and synapses represents a longstanding goal in neuroscience. Mouse connectomes could soon come within reach while human connectomes remain a more distant yet still worthy goal. Though the technologies needed to reconstruct whole-brain connectomes have not yet reached full maturity, they are advancing rapidly. Close examination of these technologies may help plan connectomics projects. Here, we quantitatively compare imaging technologies that have potential to enable whole-brain mammalian connectomics. We perform calculations on electron microscopy (EM) techniques and expansion light-sheet fluorescence microscopy (ExLSFM) methods. We consider techniques that have sufficient resolution to identify all synapses and sufficient speed to be relevant for whole mammalian brains. We offer this analysis as a resource for those considering how to organize efforts towards imaging whole-brain mammalian connectomes.

## Introduction:

Connectomics at nanoscale resolution presents a major technical challenge due to the tremendous amounts of image data involved.[1,2] To reach sufficient resolution for accurately reconstructing fine neurites and synapses, voxel dimensions on the order of just a few tens of nanometers or less are needed. But with the recent completion of the larval *Drosophila* brain connectome,[3] adult *Drosophila* brain connectome,[4,5] adult *Drosophila* ventral nerve cord connectome,[6,7] and larval *Platynereis dumerilii* connectome[8] as well as the less recent completion of the first *C. elegans* connectome,[9] it seems that the age of whole-brain connectomics has arrived. Projects with the eventual goal of mapping a whole-brain mouse connectome have recently launched, which has prompted discussions about the eventual goal of human brain connectomics. However, the *Drosophila* brain volume is only 0.0175 $mm^3$ while the mouse brain has a volume of around 500 $mm^3$ and the human brain has a volume in the range of 1,200,000 $mm^3$.[2,10,11] That is, the mouse brain has a volume of about 28,571 times that of the *Drosophila* brain, while the human brain has a volume of about 2,400 times greater still compared with the mouse brain. Dramatic increases in imaging throughput will be necessary for whole-brain mammalian connectomics endeavors.

In this analysis, we both quantitatively and qualitatively compare emerging methodologies used for imaging connectomes with a focus on the capabilities needed for the mouse brain and the human brain. The details of our calculations are available in the **Supplemental Information** (located after references). While computational image processing represents an adjacent challenge for connectomics, we here focus primarily on imaging techniques. We work under the assumption that parallel image processing developments will contribute vital new capabilities in a timely manner. In general, mammalian connectomics requires microscopy techniques featuring both high resolution and speed. Electron microscopy (EM) has thus far been the principal imaging method utilized in connectomics research because it achieves nanoscale resolution **(Figure 1A)**, though it suffers from long acquisition times even with substantial investment into ameliorating the issue.[1,12,13] Expansion microscopy coupled with light-sheet fluorescence microscopy (ExLSFM)[14–16] represents a promising alternative route that might attain similar resolution while increasing imaging speed, but this approach is still in early stages of development compared with EM. Tavakoli et al. recently showed that it is indeed possible to perform dense neuronal reconstruction using a 16-fold ExM approach and spinning-disk confocal microscopy **(Figure 1B)**,[17] but there does not yet exist a high-throughput pipeline for ExM connectomics. While the existing high-throughput EM pipeline is currently the most reliable near-term path to whole mouse brain acquisition, we would argue it is likely that an ExM-based pipeline may be developed in the relatively near future. It may be beneficial to perform rigorous comparative studies between EM and ExM pipelines (once ExM pipelines are available) to aid planning of large-scale connectomics projects. Beyond ExLSFM, ribbon scanning confocal microscopy has shown promise as an alternative fluorescence-based modality with potential for comparable performance to LSFM.[18] We do not discuss this method further since it thus far only appears in a relatively small number of publications and has not yet received attention in the context of connectomics. However, we suggest that studies combining ribbon scanning confocal microscopy with ExM may yield helpful information on the competitiveness of this technique as a connectomics method. As different imaging technologies will offer complementary capabilities for connectomics, we suggest considering carefully how to best capitalize on the advantages of each of them.

To facilitate broad comparisons across different modalities, our analysis makes a number of simplifying assumptions. Firstly, we assume voxel size is roughly proportional to resolution because many key studies do not report actual resolution (e.g. as measured by Fourier shell correlation or similar). We thus are able to use voxel size as a common metric of comparison. Since voxel size often is not directly proportional to resolution, these numbers should be thought of as best-case scenarios. We furthermore assume that isotropic voxel size of less than 18 nm or non-isotropic voxel size of less than 13×13×46 nm (<13 nm lateral, <46 nm axial) has the potential for connectomic traceability (in the case of ExM, this is effective voxel size). Several groups have used similar voxel sizes and demonstrated success with connectomic tracing.[19–21] Nonetheless, it is important to realize that traceability is largely dependent on factors such as signal-to-noise ratio, contrast, the uniformity of the sample's staining and/or expansion, the integrity of the sample's ultrastructure, and aspects of the tracing algorithms used. In EM, there is also the problem of losing or damaging a fraction of the ultrathin slices, breaking the continuity of neuronal processes.[22] We generally assume that samples have been prepared with sufficient precision to minimize damage, noise, staining issues, and section loss though we do consider that baseline levels of these factors can delineate hard limits to achievable image quality. In the section on ExLSFM, we speculatively combine a state-of-the-art technique called pan-ExM-t (which can reach 24-fold expansion)[15] with light-sheet microscopes from other publications. An exception to this is our discussion of Wang et al.'s "volumetric imaging of biological specimens via photochemical sectioning" method, where our speculation instead involves standard 4.5-fold expansion[23] to reach the assumed effective voxel size for nanoscale reconstruction since their microscope's actual voxel size was quite small. We choose light-sheet microscopes which could hypothetically achieve the aforementioned voxel sizes if their actual voxel sizes were divided by the expansion factors of 24 or 4.5. We assume that the expansion process introduces minimal distortions and minimal defects as well as that the fluorescent labeling is of sufficient quality to allow the direct application of this expansion factor. It should be noted that this assumption is not always true; ExM can introduce long-range distortions and expanded gels may experience damage during handling. Inconsistent demarcation of cellular boundaries is also an issue, though pan-ExM-t[15] and new membrane labeling methods[17,24] have substantially improved this situation. Since expanded mammalian brains may be too large for imaging as intact objects (e.g. a 24-fold expanded human brain could be over 3.5 meters in length), they will need to be sectioned either before or after expansion. As will be discussed in more depth later, difficulties in sectioning of expanded tissues could cause loss of important information as well. While we suggest that all of these issues might be conquered by sufficient optimization effort, they should still certainly be kept in mind when comparing the imaging modalities.

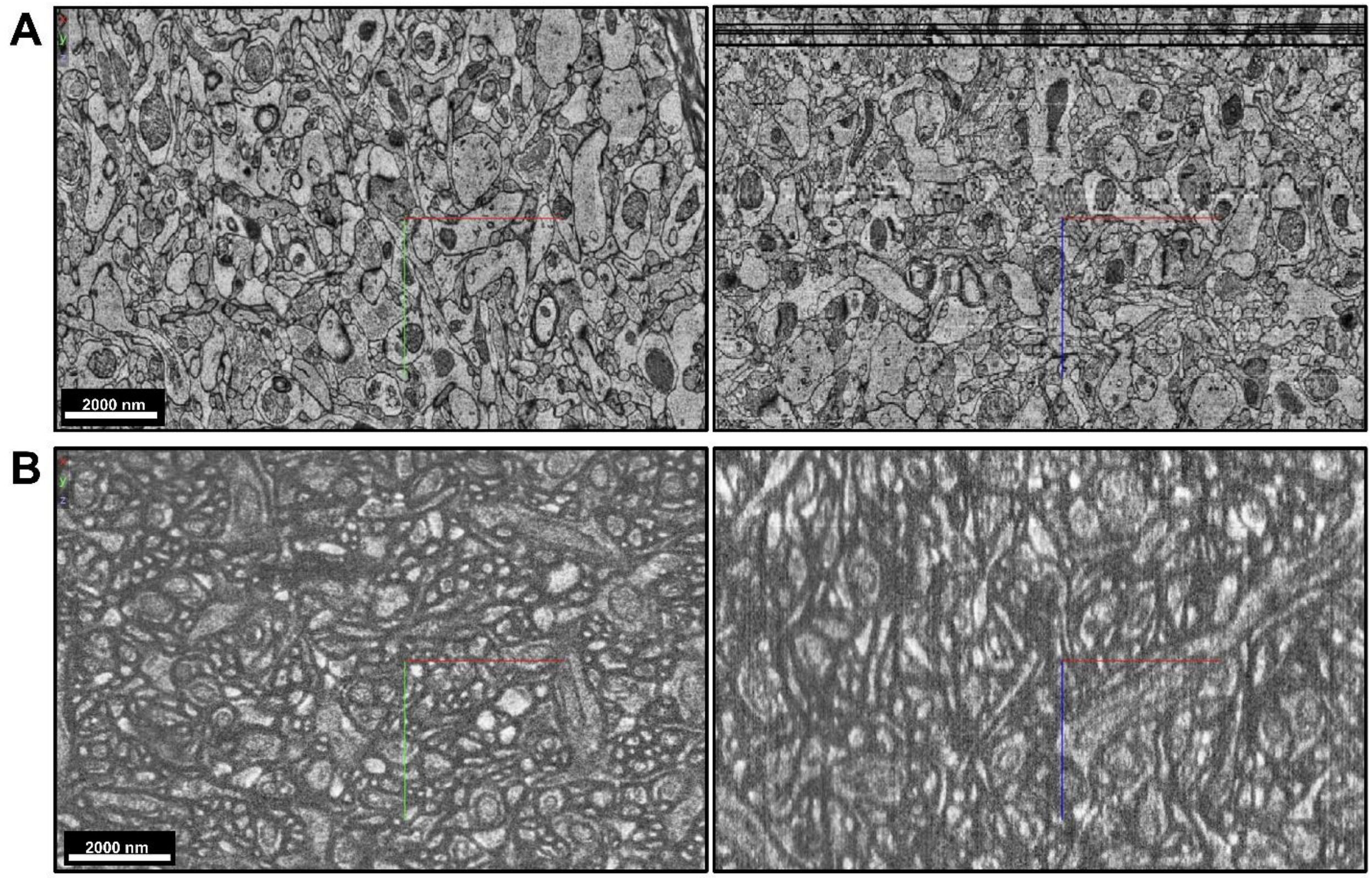

**Figure 1** Representative snapshots from the raw image data of **(A)** Shapson-Coe et al.'s ssTEM dataset of 1 $mm^3$ of human cortex[25] and **(B)** Tavakoli et al.'s dataset in mouse cortex using ExM with confocal microscopy.[17] Both of these datasets were traceable using semiautomated methods (automated segmentation plus manual correction of errors). Here, the x axes are displayed as red lines, y axes as green lines, and z axes as blue lines. We obtained these snapshots from the publicly available data within NeuroGlancer.

**An aside on computational constraints**

Though not the central focus of this analysis, computational processing of image data may also prove a limiting factor for connectomics. Some potential areas of difficulty include the transmission bandwidth of image data moving from the detector to a storage device, the massive amount of computer memory needed to store the raw image data of whole mammalian brains, compute time for stitching and aligning images, and compute time for segmentation and synapse identification.[1,26] Fully automated segmentation algorithms will furthermore be necessary for tracing the neurons and glia of whole mammalian brains and this problem has not yet been solved.[1] Contributions across many fields of research will be important to overcome these challenges. Nonetheless, we are optimistic that successes in data acquisition may drive enough interest in connectomics that the necessary computational methods will be developed.

Data storage capacity represents a particular challenge for whole-brain connectomics (note that this is distinct from RAM). Assuming isotropic voxels with dimensions of 18×18×18 nm and a memory requirement of 1 byte per voxel,[27] the raw data for a mouse brain would necessitate 85.7 petabytes and the raw data for a human brain would necessitate 205.7 exabytes of storage. As of 2023, magnetic hard disk storage cost is about $11 per terabyte,[28] which would equate to $942,700 to store the raw image data of a mouse brain or $2,262,700,000 to store the raw image data of a human brain. Mouse brain raw image data storage may thus currently be feasible, though the raw image data storage for a human brain will likely necessitate further advances in storage technology. However, historical trends[28] on data storage costs remain encouraging since consistent exponential decreases have been observed for several decades. Emerging technologies such those which use femtosecond lasers to write data into quartz glass (as seen in Microsoft's

Project Silica) might offer substantially greater storage capabilities than existing magnetic storage methods.[29] Further savings should accrue through data compression as the raw image data is processed and converted into connectome data. Indeed, a recent preprint by Li et al. demonstrated a new variational autoencoder algorithm which achieved up to 128-fold compression of EM data without significantly worsening the success of subsequent segmentation.[30] Importantly, data compression within a high-throughput pipeline will probably take place in real time as the data are collected, which will circumvent the need to store all raw uncompressed data at once. It should also be noted that data transmission during image acquisition will likely involve parallel transmission from each microscope directly into an onsite cluster, perhaps mediated by extremely high bandwidth hardware technology like InfiniBand.[31] Cloud storage might also be an option, but this could present obstacles due to generally lower bandwidth and lesser reliability.[32] Advances in hardware and software seem well on their way to handling whole mouse brain image data and may reach the point of handling whole human brain image data within the next several decades.

Image processing is another major challenge for connectomics and currently represents an even larger bottleneck than image acquisition since sufficiently accurate fully automated tracing of EM data is not yet possible[1,33] and ExLSFM still needs to mature before such issues can be addressed thoroughly.[17] Additionally, the amount of RAM necessary to efficiently run processing software on massive image datasets may be extremely high. It remains an open question as to the speed at which automated tracing software could segment image data, even assuming construction of a powerful high-performance computing (HPC) center built with the singular goal of reconstructing a mouse or human brain. It will likely be critical to leverage both application-specific hardware and innovative new tracing software strategies. As a recent example of progress on tracing software, a method by Schmidt et al. called RoboEM has demonstrated potential for more accurate automated error correction on automated connectomic segmentations.[33] By comparing to how much money was necessary for annotating past EM datasets, Schmidt et al. estimated RoboEM's computational monetary cost at 400 times less than the monetary cost of paying for manual annotation of equivalent accuracy. (Though both RoboEM and the manual annotators still made some mistakes in annotating the tested dataset, the frequency of mistakes made was roughly equivalent). RoboEM furthermore performed 3.5-fold better than state-of-the-art flood-filling networks on resolving split errors. As we will discuss more later, ExLSFM image processing requirements might drastically decrease through use of multicolor fluorescent barcodes as labels for each neuron.[14,34,35] This might provide a path forward for ExLSFM connectome image processing automation. Connectomics image processing is an area still needing further maturation.

**Electron microscopy:**

EM represents a central method used in connectomics studies and has so far been the only technique that has facilitated computational reconstruction of complete whole-brain connectomes (*C. elegans*, *D. melanogaster*, and *P. dumerilii*).[3–5,8,36] It involves sectioning (or milling) a sample into very thin slices and sequentially imaging each slice with an electron microscope. The most commonly used EM methods in connectomics are serial-section transmission electron microscopy (ssTEM) **(Figure 2A)**, serial block-face electron microscopy (SBEM) **(Figure 2B)**, and focused ion beam scanning electron microscopy (FIB-SEM) **(Figure 2C)**.[37] Extensive engineering efforts have gone toward increasing the throughput of these core techniques. Some important advances in the area include multibeam scanning electron microscopy (multiSEM) **(Figure 2D)**,[27] automated tape-collecting ultramicrotomy scanning electron microscopy (ATUM-SEM)[38] **(Figure 2E)**, GridTape ssTEM[39] **(Figure 2F)**, and gas cluster ion beam scanning electron microscopy (GCIB-SEM)[40] **(Figure 2G)**. These approaches have shown success for small volumes (i.e. up to the 1 $mm^3$ scale) and may be amenable to parallelization for higher throughput, albeit at very high monetary costs. Indeed, some investigators have suggested industrial facilities with hundreds of electron microscopes working in parallel to map entire mammalian brains.[40] Constructing and maintaining such facilities would probably cost hundreds of millions to billions of dollars, so further increases in the throughput of individual instruments may still be necessary. Nonetheless, EM brings a long history of connectomics successes and scaling it up for the mouse brain connectome has been suggested as feasible within the coming decade.[2]

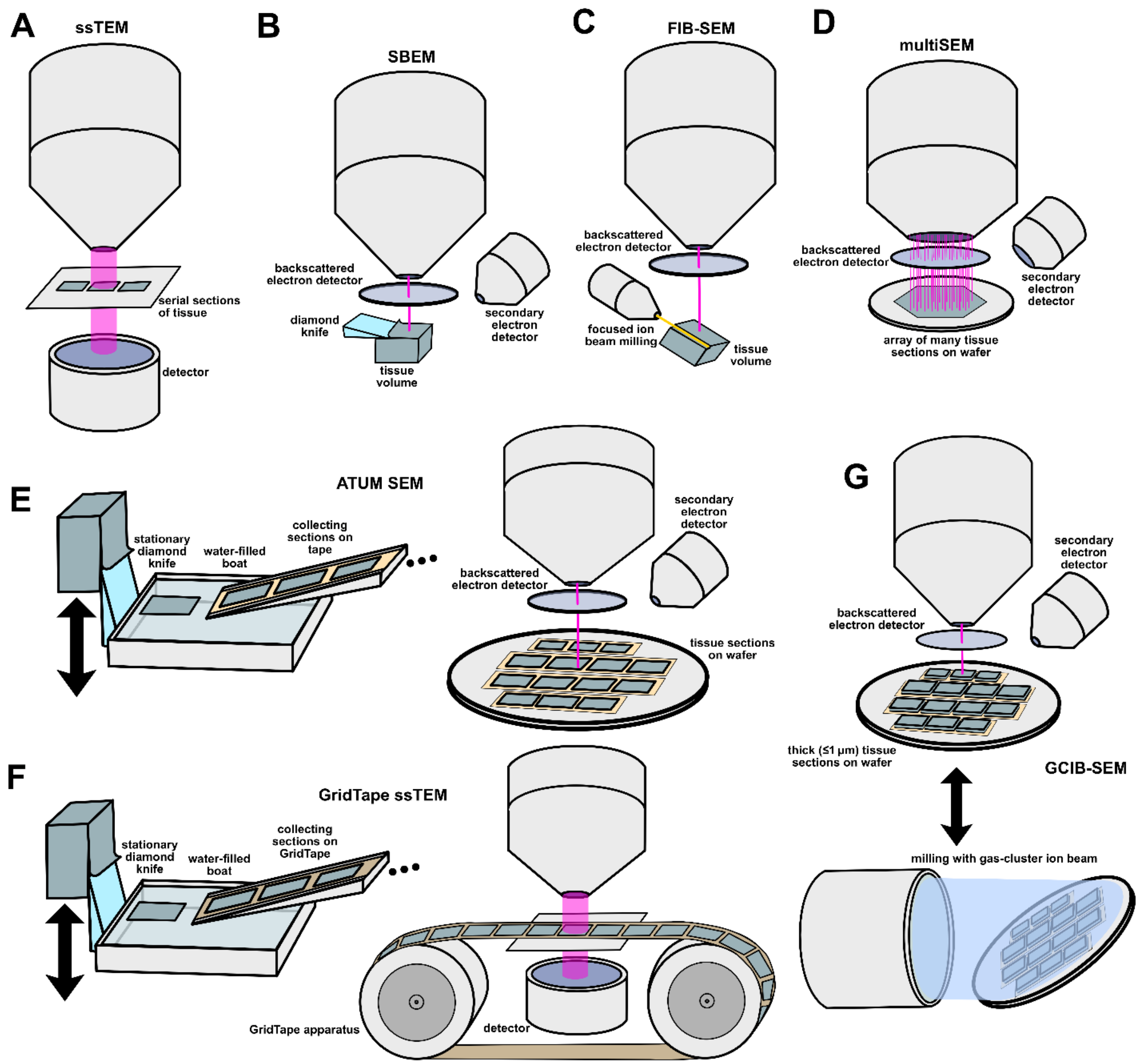

**Figure 2** Depictions of the mechanisms of **(A)** Serial-section transmission electron microscopy (ssTEM) **(B)** serial block-face scanning electron microscopy (SBEM), **(C)** Focused ion beam scanning electron microscopy (FIB-SEM), **(D)** multibeam scanning electron microscopy (multiSEM), **(E)** automated tape-collecting ultramicrotomy scanning electron microscopy (ATUM-SEM), **(F)** GridTape ssTEM, and **(G)** gas cluster ion beam scanning electron microscopy (GCIB-SEM). For more details on these technologies, their use in connectomics, and how they have been modified for improved throughput, see excellent reviews by Titze et al.[38] and by Peddie et al.[41]

How fast are current EM methods for imaging brain tissue? **(Table 1)** In a landmark study by Yin et al., a 1 $mm^3$ volume of brain tissue was imaged in 6 months (4×4×40 nm voxels) using a system consisting of 6 parallelized and fully automated ssTEM microscopes.[13] The authors built this system by incorporating several custom hardware modifications into commercially available JEOL 1200EXII 120 kV TEMs as well as by developing a software framework to facilitate a consistent automation process. Each microscope cost about $125K and the hardware modifications on each added a further $125K to the cost ($250K per customized microscope). As such, the total cost of the system of 6 microscopes was about $1.5M. While these microscopes were purchased secondhand and thus were somewhat less expensive than brand new instruments would have been, economies of scale may act to counterbalance the price of such microscopes in the future. Another example of a 1 $mm^3$ acquisition comes from Shapson-Coe et al. where ATUM was used to section the tissue and a multiSEM instrument (4×4×33 nm voxels) was used to image the tissue over the course of 326 days (10.516 months).[25,42] The price of the multiSEM instrument was about $4M for the version used by Shapson-Coe et al., which utilized 61 beams.[25,39,42] It should be noted that an even faster 91-beam multiSEM is also now commercially available. These investigations represent the first steps into millimeter-scale EM connectomics and will serve as a foundation for future advances in the field.

EM throughput may continue to grow over time as new technologies are developed. A promising direction employs the techniques of broad ion beam scanning electron microscopy (BIB-SEM) and multi-beam scanning transmission electron microscopy (mSTEM) and combines them to make a method called BIB-mSTEM.[43] This involves cutting thick sections of around 100 nm to 10,000 nm, imaging them via mSTEM, milling via BIB, and computationally stitching reconstructions of the thick sections. Thick sections are much easier to handle and cause less difficulty in computational reconstruction downstream than ultrathin 30-40 nm sections, so the technique has potential to accelerate connectomics workflows by circumventing the slowdowns associated with ultrathin sectioning difficulties. Zheng et al. also greatly improved a technique called beam deflection TEM (bdTEM), which has been shown to substantially increases the throughput relative to the Yin et al. study described earlier.[44] Beam deflection allows scanning of the electron beam over nine image tiles without moving the stage, thus eliminating eight out of nine stage movements and speeding up imaging overall. With the same type of camera, the bdTEM microscopes can each image tissue (3.6×3.6×45 nm voxels) much faster than those of the Yin et al. study. The authors estimate that if the four microscopes were to run at 65% uptime, they could together image 1 $mm^3$ in 37 days. Also retrofitted from JEOL 1200EXII 120 kV TEMs, each of the custom bdTEMs cost about $500K. Four of these were constructed at the Princeton facility described by Zheng et al. (about $2M total cost). Future innovations have potential to continue improving the throughput of EM instruments.

**Table 1** Comparison of parameters from key publications and preprints involving high-throughput EM methods for connectomics.

| **Study** | **Yin et al.[13] (2020)** | **Shapson-Coe et al.[25] (2024)** | **Zheng et al.[44] (2024)** |
|---|---|---|---|
| **Imaging speed per microscope ($mm^3$/month)** | 0.0278 | 0.0951 | 0.3 (projected) |
| **Cost per microscope** | $250,000 | $4M | $500,000 |
| **Voxel size (nm)** | 4×4×40 | 4×4×33 | 3.6×3.6×45 |
| **Time to image mouse brain per 1 microscope** | 1,500 years | 438 years | 139.6 years |
| **Time to image human brain per 1 microscope** | 3.6M years | 1.05M years | 333,333 years |
| **Time to image mouse brain with $100M of microscopes** | 3.75 years with 400 microscopes | 17.5 years with 25 microscopes | 0.69 years with 200 microscopes |
| **Time to image human brain with $100M of microscopes** | 9,000 years with 400 microscopes | 42,061 years with 25 microscopes | 1,667 years with 200 microscopes |

Although we assume in our calculations that EM sample preparation difficulties can be overcome during the design of high-throughput imaging pipelines, this part of the process remains challenging. Serial sections can suffer from inconsistent staining, warping, folding, shrinkage, and ultrastructural damage.[41,45] Problems with staining have been greatly ameliorated by improved en bloc staining techniques where large tissue volumes and even whole mouse brains can be consistently penetrated by the reagents.[46] To accelerate sectioning and to improve the quality and consistency of the resulting sections, the aforementioned automated tape-collecting ultramicrotome (ATUM) was developed.[47] ATUM originally used electron-opaque support tape, limiting its use to SEM devices.[48] More recently, an approach known as GridTape has made ATUM compatible with TEM.[39] GridTape utilizes an electron lucent type of support tape for ATUM. It thus combines automated serial sectioning onto a TEM-compatible tape substrate with automated high-throughput imaging, creating a fully automated workflow for ssTEM. It has shown the capacity to collect over 4,000 sections (each roughly 1 $mm^2$ in area) per day. At 40 nm thickness per section, this means that a ~0.16 $mm^3$ volume (roughly 1 $mm^2$ area times 0.16 mm thickness) could be prepared in a day using one GridTape instrument. With the exact setup described in the original GridTape publication,[39] sample preparation would still take a large amount of time (e.g. 8.56 years for one GridTape instrument working on a 500 $mm^3$ mouse brain). However, alternative designs might greatly speed up this process. A wider diamond knife may allow preparation of sections with substantially larger surface area. For instance, reliable cutting of ATUM sections of 3×3 $mm^2$ using a 4 mm blade have been reported and even larger blades may further increase throughput on a single instrument.[38] Cameras with wider field of view (FOV) are available for such purposes as well. Similar reasoning could apply to SBEM methods. FIB-SEM methods can also cause unintended damage to the tissue during ion beam milling and the likelihood of inconsistent milling increases for wider tissue volumes since the energy of the beam decreases as it travels through tissue for long distances.[41] However, GCIB-SEM may circumvent many of these issues by replacing the parallel milling beam with a perpendicular beam (of lower energy) to allow consistent milling across arbitrarily wide surfaces.[40] Because of these factors, we suggest that imaging rate may still represent the main speed bottleneck in ssTEM, SBEM, and GCIB-SEM connectomics

**Light-sheet fluorescence microscopy with expansion microscopy:**

Coupling LSFM with ExM represents a promising up-and-coming approach for mammalian brain connectomics. LSFM utilizes specialized lenses to generate a thin sheet of laser light and excite fluorophores across a planar region of tissue.[49] Alternatively, LSFM can utilize a rapid laser scanning approach to construct a sheet of light along the axis of the scan. In either case, an objective lens to collect light for the detector is placed perpendicular to the light sheet. The instrument translates the sheet (or the sample itself) along this perpendicular axis to collect a stack of images for reconstruction into a 3D volume. For most samples, LSFM requires tissue clearing to minimize scattering of the light. Fortunately, ExM clears tissue by virtue of volumetric dilution, so expanded tissues are translucent and amenable to LSFM.[50] ExLSFM is a serendipitous union of technologies, combining increased resolution with the speed necessary to reconstruct large volumes.

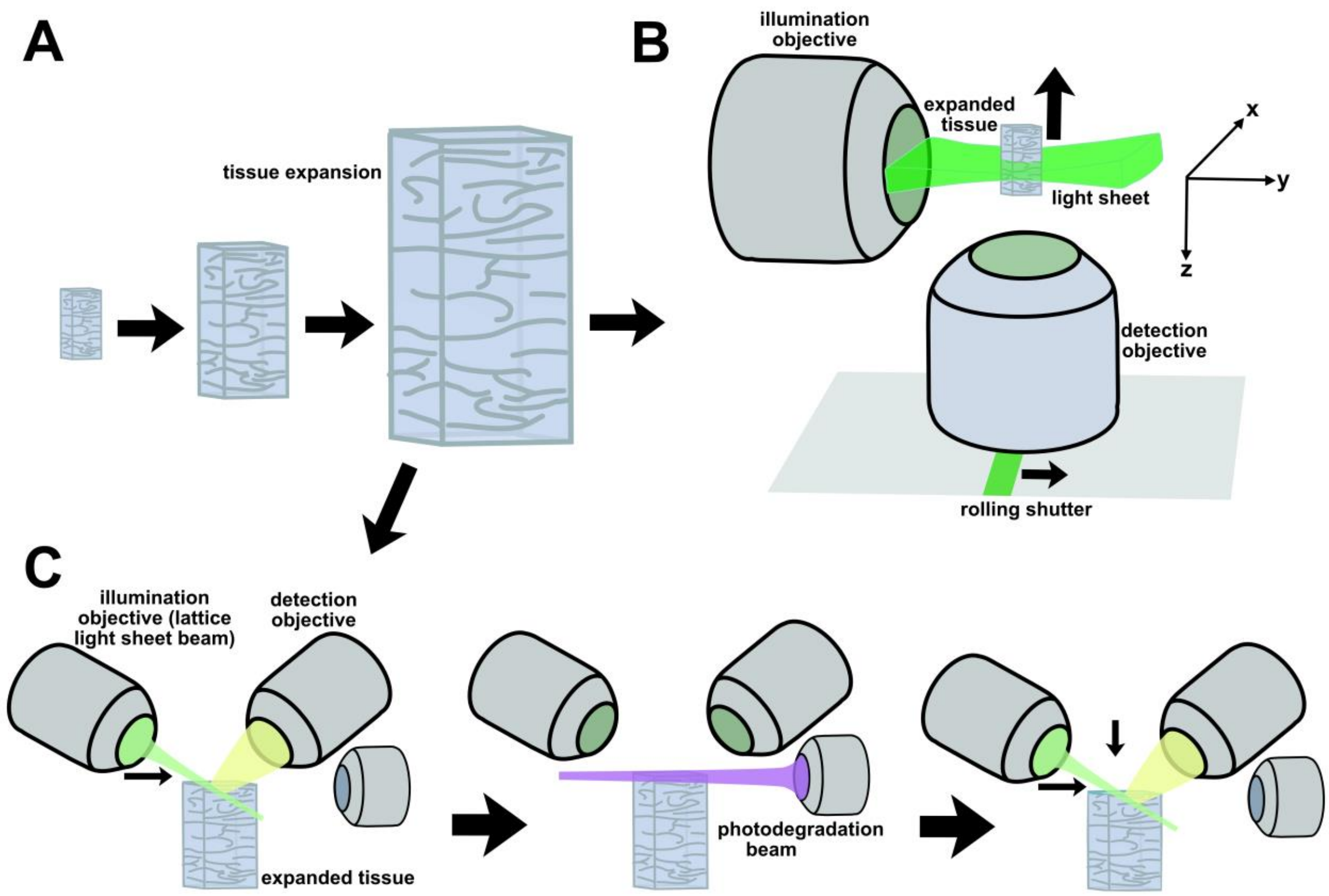

**Figure 3 (A)** Tissue expansion physically enlarges tissue after infusion with a swellable hydrogel and addition of purified water. More details on expansion microscopy (ExM) can be found in excellent reviews by Truckenbrodt[50] and Wen et al.[51] **(B)** Expansion light-sheet fluorescence microscopy (ExLSFM) in an axially swept light-sheet microscopy (ASLM) configuration. **(C)** Expansion lattice light-sheet microscopy (ExLLSM) in a "volumetric imaging of biological specimens via photochemical sectioning" (VIPS) configuration. More details on light-sheet fluorescence microscopy (LSFM) can be found in an excellent review by Stelzer et al.[49]

Although ExLSFM has not yet reached a point where it is widely used in dense neural circuit mapping, it represents the subject of intensive research and seems on the cusp of feasibility as a tool for connectomics **(Table 2)**. As mentioned earlier, Tavakoli et al. successfully achieved dense neuronal reconstruction of 16-fold expanded tissue imaged with a spinning-disk confocal microscope.[17] They imaged a 16-fold expanded piece of tissue (with a pre-expansion volume of 0.00095 $mm^3$) over the course of 6.5 hours. This is too slow for mammalian brain connectomics since 1 $mm^3$ would take 6,841.9 hours or 9.37 months. It is plausible that applying LSFM instead of spinning-disk confocal microscopy could greatly accelerate this type of imaging. However, since their spinning-disk confocal setup had much better axial and lateral resolution than is achievable by the faster types of LSFM (discussed in following paragraphs), the expansion factor would likely need to be somewhat higher to still facilitate dense reconstruction. With respect to this point, another method by M'Saad et al. called pan-ExM of tissue (pan-ExM-t) has achieved 24-fold expansion as well as clear enough staining of proteins to delineate cellular boundaries in a fashion that somewhat resembles EM.[15] While M'Saad et al.'s expansion technique was not used in combination with LSFM, it seems plausible that this will be a next step. Furthermore, in the case of lattice light-sheet microscopy (LLSM), a standard 4.5-fold expansion factor[23] might be sufficient. A recently developed staining technique by Shin et al. has also shown the capacity to strongly label membranes in expanded and iteratively expanded brain tissue samples.[24] This staining technique has high potential to improve the traceability of brain tissue in the context of ExM. Along these lines, though volumetric signal dilution from high expansion factors sometimes necessitates longer exposure times for successful imaging, dense staining approaches like the aforementioned pan-labeling technique by M'Saad et al.[15] and membrane labeling method by Shin et al.[24] have potential to counteract such slowdowns. Iterative signal amplification methods[52–54] or extremely bright plasmonic fluorophore labels[55,56] may further improve staining quality in expanded tissues and thus ensure that exposure requirements do not substantially slow imaging. As will be discussed subsequently, certain LSFM setups can image at sufficiently small voxel sizes that 24-fold expansion or even 4.5-fold expansion perhaps along with membrane labeling have potential to produce EM-like resolution.

What kinds of ExLSFM will be useful for imaging mammalian connectomes? **(Figure 3A-C)** LLSM has received a great deal of attention,[57–59] but it usually suffers from extremely small FOV and working distance limitations which may make it unsuitable for whole-brain mammalian connectomics, especially in the context of expanded samples.[60,61] (Small FOV vastly slows imaging due to volumetric scaling). However, Wang et al. developed a promising workaround in a recent study on ExM with LLSM (ExLLSM): a photodegradable hydrogel and a UV light sheet for photochemical sectioning **(Figure 3C)**.[62] In this way, large tissue volumes were rapidly reconstructed through sequential ExLLSM imaging followed by controlled photodegradation of the imaged layers, effectively extending the "working distance" of the ExLLSM. They named this method volumetric imaging of biological specimens via photochemical sectioning (VIPS). Wang et al. demonstrated the potential of their technology by imaging a mouse olfactory bulb of 9.4×6.4×4.3 mm (258.7 $mm^3$) volume after 2-fold expansion in 10 days at an actual voxel size of 56×56×135 nm (28×28×67.5 nm effective voxel size post-expansion). Although successful for reconstructing axons, this was insufficient for reconstruction of some dendrites and synapses. As such, we will assume that 4.5-fold expansion[63] would be sufficient for full nanoscale connectome reconstruction since this reaches 12.4×12.4×30 nm non-isotropic voxel size (see introduction for more details on our assumptions). With 4.5-fold expansion instead of 2-fold expansion, the same volume of 258.7 $mm^3$ should take about 113.9 days to image (since the volume would be about 11.39 times larger). At this imaging speed, VIPS has promise for whole-brain mammalian connectomics, particularly if the method can be improved further. Another valuable approach may come in the form of axially swept light-sheet microscopy (ASLM) **(Figure 3B)**. In particular, Chakraborty et al. demonstrated cleared-tissue axially swept light-sheet microscopy (ctASLM), which was able to image 1 $mm^3$ volumes at 425 nm isotropic step size (voxel size) over the course of 17.6 minutes.[64] If coupled with 24-fold expansion from pan-ExM-t, this would result in effective 17.7 nm isotropic voxels and would take 13824 times longer to image (about 169 days) due to the larger volume. With an extension of the ctASLM design called signal improved ultrafast light-sheet microscopy (SIFT), Prince et al. were able to improve the imaging speed to roughly 1 $mm^3$ per 9.4 minutes while maintaining 425 nm voxels.[65] Once again, if coupled with 24-fold expansion from pan-ExM-t, this would result in effective 17.7 nm isotropic voxels and would once again take 13,824 times longer to image (about 90 days) due to the volume increase. It seems that the dawn of scalable ExLSFM connectomics may arrive soon given the ongoing convergence of rapidly advancing ExM and LSFM technologies.

We should also mention a highly publicized advance wherein a recent Allen Institute effort constructed a LSFM microscope called ExA-SPIM (expansion-assisted selective plane illumination microscopy), which is specially designed to rapidly image large volumes of expanded tissue at moderate voxel sizes of at best 1,000×1,000×2,500 nm.[66] Though the ExA-SPIM is very fast due to its massive FOV and working distance and it can reach an imaging rate of 50 $cm^3$ per day, even 24-fold expansion would only give effective voxel size of 41.67×41.67×104.17 nm. This is likely insufficient to achieve the resolution necessary for nanoscale connectomics. Because of this drawback, we will not include ExA-SPIM in our tabular comparison of ExLSFM methods for whole-brain mammalian connectomics.

Barcoding neurons by expressing unique mixtures of fluorophores or by using multi-round imaging of sequences of distinct fluorophores has been proposed as a way to mitigate the need for extremely high resolution in connectomics.[14,34,35] Feasibility of dense connectomics through a combination of Brainbow and sequential barcode methods has been supported by computational modeling on simulated data.[14,34] However, the literature is lacking in quantitative metrics as to exactly how much these contrast mechanisms might mitigate the need for high resolution. We suggest prioritizing further quantitative investigations exploring the degree that color contrast and barcodes may allow for connectomics imaging at coarser voxel size. It should also be noted that sequential barcodes multiply the acquisition time since they generally necessitate multiple rounds of imaging, though this is fortunately a linear increase rather than an exponential one. Regardless of its effect on imaging time, adding color contrast and neuronal barcodes will likely improve segmentation accuracy, particularly since existing single-color datasets are still subject to time-consuming manual proofreading steps.[4,25]

**Table 2** Comparison of parameters from key publications and preprints involving high-throughput LSFM methods as hypothetically combined with either standard 4.5-fold expansion[23] or 24-fold pan-ExM-t expansion as described by M'Saad et al.'s preprint[15] and also possibly with the membrane stain from Shin et al.'s preprint.[24]

| **Hypothetical combined study** | **Wang et al.[62] (2024) with 4.5-fold expansion** | **Chakraborty et al.[64] (2019) with 24-fold expansion[15]** | **Prince et al.[65] (2024) with 24-fold expansion[15]** |
|---|---|---|---|
| **Actual imaging speed per microscope ($mm^3$/month)** | 802 | 2,536 | 3,707 |
| **Cost per microscope** | $300,000 | $120,000 | $104,000 |
| **Effective voxel size (nm) after expansion** | 12.4×12.4×30 | 17.7×17.7×17.7 | 17.7×17.7×17.7 |
| **Time to image expanded mouse brain per 1 microscope** | 4.7 years | 227 years | 155 years |
| **Time to image expanded human brain per 1 microscope** | 11,363 years | 545,000 years | 373,000 years |
| **Time to image expanded mouse brain with $100M of microscopes** | 0.014 years with 333 microscopes | 0.273 years with 833 microscopes | 0.162 years with 962 microscopes |
| **Time to image expanded human brain with $100M of microscopes** | 34 years with 333 microscopes | 654 years with 833 microscopes | 388 years with 962 microscopes |

While we largely assume in our calculations that ExLSFM sample preparation difficulties can be overcome over the course of the design of high-throughput imaging pipelines, this part of the process may present unique challenges as well. In particular, there has been little work towards developing reliable strategies for either pre-expansion or post-expansion tissue sectioning. As mentioned in the introduction, the large size of a whole expanded brain will necessitate sectioning to fit within LSFM FOVs, which could destroy parts of the tissue. Furthermore, a way of holding the tissue steady during slicing will be needed to ensure clean cuts. But in the pre-expansion case, traditional forms of embedding may prove problematic since they could preclude subsequent expansion. That said, it is conceivable one might develop a sturdy yet under certain conditions removable embedding medium to facilitate pre-expansion sectioning. For post-expansion sectioning, one could embed the expanded brain tissue, but the physical properties of the expansion gel might make clean cuts difficult to achieve. Laser-based cutting methods may be useable in this case since, unlike a microtome blade, they would not be prone to problems arising from the gel's mechanical elasticity. The fine details of the neurites in expanded tissue would be physically larger, therefore the part of the gel ablated by the laser may contain fewer cellular features than if the laser had been applied to a non-expanded sample. Indeed, the techniques of the earlier mentioned VIPS approach[62] might be adapted for this purpose, though they would need optimization to slice tissue rather than fully ablate. Some degree of work remains to make these possibilities a reality. We argue that once sectioning methods are established, they will contribute relatively little to the overall costs of connectome projects because a vastly smaller number of cuts will need to be made to each brain compared to the number of cuts needed for EM. Nonetheless, efficacious sectioning methods will remain crucially important to the success of whole-brain ExLSFM connectomics efforts.

**Discussion:**

Making comparisons across the landscape of connectome imaging technologies is difficult because distinct methods involve different technical challenges. Furthermore, some degree of extrapolation is necessary to account for the rapid development of emerging but not yet fully tested imaging pipelines. Nonetheless, careful examination of existing and emerging technologies can yield useful insights **(Table 3)**. EM and ExLSFM have their own unique pros and cons **(Table 4)**. EM benefits from a long history of technology development for connectomics applications and has a strong existing infrastructure for high-throughput connectomics pipelines. ExLSFM represents an emerging modality that has only recently been considered for connectomics, yet it has a speed advantage over EM and opens doors to using multiplexed fluorescent markers for simultaneously probing the brain's molecular information.[14] Each of these technologies deserves consideration in the context of mammalian connectomics, though developing multiple imaging modalities in parallel will likely prove the best route in the end.

Imaging mammalian connectomes represents a grand challenge. Despite this, new enabling technologies are rapidly maturing. *C. elegans* and *Drosophila* connectomes have already transformed the practice of neurobiology and led to a wide variety of advances in the field. Coupled with computational modeling, mammalian connectomes could provide unprecedented insights into cognition, emotion, and brain disease. They may inform the design of AI, robotics, and brain-computer interfaces.[67] Advanced imaging technologies are drawing mouse connectomes within reach. Mapping the human connectome remains a more distant goal, yet paradigm-shifting advances such as synchrotron x-ray nanotomography methods[68–70] and/or massive parallelization may still make that a feasible goal sometime in the future. We anticipate that mammalian connectomes will facilitate dramatically transformative insights, improving biomedicine and perhaps even helping us better understand what it means to be human.

**Table 3** Direct comparisons of important characteristics of leading EM and ExLSFM approaches.

| | **EM** | **ExLSFM** |
|---|---|---|
| **Time to image mouse brain with $100M of fastest reported instruments** | 0.69 years with 200 bdTEM instruments at voxel size of 3.6×3.6×45 nm[44] | 0.014 years with 333 VIPS ExLLSM instruments at effective voxel size of 12.4×12.4×30 nm[62] |
| **Time to image human brain** | 1,667 years with 200 bdTEM instruments at voxel size of 3.6×3.6×45 nm[44] | 34 years with 333 VIPS ExLLSM instruments at effective voxel size of 12.4×12.4×30 nm[62] |
| **with $100M of fastest reported instruments** | | |
| **Sample preparation time and challenges** | A single ATUM instrument with a 4 mm blade may section around 3×3×0.16 mm/day,[38,39] so 200 ATUMs could section a whole mouse brain in 1.74 days or a whole human brain in 11.42 years. | ExM protocols on their own often take <1 week.[15,71] However, brains will likely need to be cut into millimeter-scale or centimeter-scale blocks and there is currently no method for properly sectioning pre-expansion or post-expansion tissues. |
| **Raw data storage requirements** | ~85.7 petabytes (cost of roughly $942,700) for mouse brain raw data and ~205.7 exabytes (cost of roughly $2,262,700,000) for human brain raw data.[28] | Identical raw data storage requirements *unless* fluorescent color contrast methods (e.g. Brainbow[72]) or fluorescent barcoding[14,34] enable acquisition of larger voxels without loss of traceability. It should be noted that the number of color channels scales linearly while larger voxels could exponentially decrease necessary data storage amounts. |
| **Data processing requirements** | Much work remains to develop fully automated tracing algorithms. As there are no existing examples of fully automated tracing software, we cannot estimate compute requirements. Advances in application-specific hardware may mitigate HPC costs. | Molecular labels in the form of fluorescent color contrast or fluorescent barcoding might help partially mollify the obstacles of automated tracing, but these approaches are still in their infancy. Additionally, alignment of multiple expanded volumes (after sectioning) may present new challenges. |
| **Feasibility of mouse connectome** | An EM mouse brain connectome project is feasible in the next decade or so if the fully automated tracing problem can be solved and sufficient funding is available to pay for microscopes, data storage and processing, labor costs, etc. (on the order of at least several hundred million dollars).[1,2] | An ExLSFM mouse brain connectome project is feasible in the next decade or so if ExM sample preparation and imaging pipelines mature. Lack of approaches for sectioning expanded tissue volumes and reconstructing across sections hampers progress for now. |
| **Feasibility of human connectome** | EM-based reconstruction of a human connectome is unlikely to occur for a very long time. Alternative technological paradigms like x-ray nanotomography[68–70] should be explored prior to making massive investments into a human connectome project. | An ExLFSM human brain connectome project might be feasible in ~25-50 years assuming further technological advances in both ExM and LSFM, but this remains highly uncertain. New technological paradigms such as x-ray nanotomography[68–70] should be explored before making enormous investments into a human connectome project. |

**Table 4** Brief overview of some advantages and some disadvantages of EM and ExLSFM.

| | **Advantages** | **Disadvantages** |
|---|---|---|
| **EM:** | • Well-established technology<br>• Track record for invertebrate connectomes | • Long acquisition times even with advances<br>• Involves lots of destructive slicing or milling steps |
| | • High resolution<br>• Established membrane staining methods<br>• Can reimage desired subvolumes at higher magnifications to uncover more details | • Not easily compatible with molecular labeling<br>• MultiSEM instruments are very expensive |
| **ExLSFM** | • Much higher speed<br>• Compatible with molecular labeling<br>• Compatible with barcoding<br>• Less slicing needed | • Less well-established technology<br>• Lack of connectomics track record<br>• Less well-established membrane staining methods<br>• Imperfect expansion might damage cellular features |

**Acknowledgements:**
We thank Dr. Anders Sandberg for engaging with us in helpful discussions that helped improve this paper.

**Declaration of competing interests:**
The authors declare no competing interests.

**References:**
1. Motta, A., Schurr, M., Staffler, B. & Helmstaedter, M. Big data in nanoscale connectomics, and the greed for training labels. *Curr. Opin. Neurobiol.* **55**, 180–187 (2019).
2. Abbott, L. F. *et al.* The Mind of a Mouse. *Cell* **182**, 1372–1376 (2020).
3. Winding, M. *et al.* The connectome of an insect brain. *Science (80-. ).* **379**, eadd9330 (2023).
4. Dorkenwald, S. *et al.* Neuronal wiring diagram of an adult brain. *Nature* **634**, 124–138 (2024).
5. Schlegel, P. *et al.* Whole-brain annotation and multi-connectome cell typing of Drosophila. *Nature* **634**, 139–152 (2024).
6. Takemura, S. *et al.* A Connectome of the Male Drosophila Ventral Nerve Cord. at https://doi.org/10.7554/elife.97769.1 (2024).
7. Azevedo, A. *et al.* Connectomic reconstruction of a female Drosophila ventral nerve cord. *Nature* **631**, 360–368 (2024).
8. Verasztó, C. *et al.* Whole-body connectome of a segmented annelid larva. at https://doi.org/10.7554/elife.97964.1 (2024).
9. White, J. G., Southgate, E., Thomson, J. N. & Brenner, S. The structure of the nervous system of the nematode Caenorhabditis elegans. *Philos Trans R Soc L. B Biol Sci* **314**, 1–340 (1986).
10. Cosgrove, K. P., Mazure, C. M. & Staley, J. K. Evolving Knowledge of Sex Differences in Brain Structure, Function, and Chemistry. *Biol. Psychiatry* **62**, 847–855 (2007).
11. Allen, J. S., Damasio, H. & Grabowski, T. J. Normal neuroanatomical variation in the human brain: An MRI-volumetric study. *Am. J. Phys. Anthropol.* **118**, 341–358 (2002).
12. Mikula, S. Progress Towards Mammalian Whole-Brain Cellular Connectomics. *Frontiers in Neuroanatomy* vol. 10 62 at https://www.frontiersin.org/article/10.3389/fnana.2016.00062 (2016).
13. Yin, W. *et al.* A petascale automated imaging pipeline for mapping neuronal circuits with high-throughput transmission electron microscopy. *Nat. Commun.* **11**, 4949 (2020).
14. Yoon, Y.-G. *et al.* Feasibility of 3D Reconstruction of Neural Morphology Using Expansion Microscopy and Barcode-Guided Agglomeration . *Frontiers in Computational Neuroscience* vol. 11 at https://www.frontiersin.org/articles/10.3389/fncom.2017.00097 (2017).
15. M'Saad, O. *et al.* All-optical visualization of specific molecules in the ultrastructural context of brain tissue. *bioRxiv* 2022.04.04.486901 (2022) doi:10.1101/2022.04.04.486901.
16. Lillvis, J. L. *et al.* Rapid reconstruction of neural circuits using tissue expansion and light sheet microscopy. *Elife* **11**, e81248 (2022).

17. Tavakoli, M. R. *et al.* Light-microscopy based dense connectomic reconstruction of mammalian brain tissue. *bioRxiv* 2024.03.01.582884 (2024) doi:10.1101/2024.03.01.582884.
18. Watson, A. M. *et al.* Ribbon scanning confocal for high-speed high-resolution volume imaging of brain. *PLoS One* **12**, e0180486 (2017).
19. Motta, A. *et al.* Dense connectomic reconstruction in layer 4 of the somatosensory cortex. *Science (80-. ).* **366**, eaay3134 (2019).
20. Schmidt, H. *et al.* Axonal synapse sorting in medial entorhinal cortex. *Nature* **549**, 469–475 (2017).
21. Xu, C. S., Pang, S., Hayworth, K. J. & Hess, H. F. Enabling FIB-SEM Systems for Large Volume Connectomics and Cell Biology. *bioRxiv* 852863 (2019) doi:10.1101/852863.
22. Lee, K. *et al.* Convolutional nets for reconstructing neural circuits from brain images acquired by serial section electron microscopy. *Curr. Opin. Neurobiol.* **55**, 188–198 (2019).
23. Asano, S. M. *et al.* Expansion Microscopy: Protocols for Imaging Proteins and RNA in Cells and Tissues. *Curr. Protoc. Cell Biol.* **80**, e56 (2018).
24. Shin, T. W. *et al.* Dense, Continuous Membrane Labeling and Expansion Microscopy Visualization of Ultrastructure in Tissues. *bioRxiv* 2024.03.07.583776 (2024) doi:10.1101/2024.03.07.583776.
25. Shapson-Coe, A. *et al.* A petavoxel fragment of human cerebral cortex reconstructed at nanoscale resolution. *Science (80-. ).* **384**, eadk4858 (2024).
26. Beyer, J. *et al.* A Survey of Visualization and Analysis in High-Resolution Connectomics. *Comput. Graph. Forum* **41**, 573–607 (2022).
27. Eberle, A. L. & Zeidler, D. Multi-Beam Scanning Electron Microscopy for High-Throughput Imaging in Connectomics Research. *Frontiers in Neuroanatomy* vol. 12 112 at https://doi.org/10.3389/fnana.2018.00112 (2018).
28. Our World in Data: Historical price of computer memory and storage. https://ourworldindata.org/grapher/historical-cost-of-computer-memory-and-storage (2024).
29. Anderson, P. *et al.* Project Silica: Towards Sustainable Cloud Archival Storage in Glass. in *Proceedings of the 29th Symposium on Operating Systems Principles* 166–181 (Association for Computing Machinery, 2023). doi:10.1145/3600006.3613208.
30. Li, Y. *et al.* EM-Compressor: Electron Microscopy Image Compression in Connectomics with Variational Autoencoders. *bioRxiv* 2024.07.07.601368 (2024) doi:10.1101/2024.07.07.601368.
31. Lu, P.-J., Lai, M.-C. & Chang, J.-S. A Survey of High-Performance Interconnection Networks in High-Performance Computer Systems. *Electronics* vol. 11 at https://doi.org/10.3390/electronics11091369 (2022).
32. Tahir, A. *et al.* A Systematic Review on Cloud Storage Mechanisms Concerning e-Healthcare Systems. *Sensors* vol. 20 at https://doi.org/10.3390/s20185392 (2020).
33. Schmidt, M., Motta, A., Sievers, M. & Helmstaedter, M. RoboEM: automated 3D flight tracing for synaptic-resolution connectomics. *Nat. Methods* **21**, 908–913 (2024).
34. Chen, S., Loper, J., Zhou, P. & Paninski, L. Blind demixing methods for recovering dense neuronal morphology from barcode imaging data. *PLOS Comput. Biol.* **18**, e1009991 (2022).
35. Shen, F. Y. *et al.* Light microscopy based approach for mapping connectivity with molecular specificity. *Nat. Commun.* **11**, 4632 (2020).
36. Cook, S. J. *et al.* Whole-animal connectomes of both Caenorhabditis elegans sexes. *Nature* **571**, 63–71 (2019).
37. Kubota, Y., Sohn, J. & Kawaguchi, Y. Large volume electron microscopy and neural microcircuit analysis. *Front. Neural Circuits* **12**, 98 (2018).
38. Titze, B. & Genoud, C. Volume scanning electron microscopy for imaging biological ultrastructure. *Biol. Cell* **108**, 307–323 (2016).
39. Graham, B. J. *et al.* High-throughput transmission electron microscopy with automated serial sectioning. *bioRxiv* 657346 (2019) doi:10.1101/657346.
40. Hayworth, K. J. *et al.* Gas cluster ion beam SEM for imaging of large tissue samples with 10 nm

isotropic resolution. *Nat. Methods* (2019) doi:10.1038/s41592-019-0641-2.
41. Peddie, C. J. *et al.* Volume electron microscopy. *Nat. Rev. Methods Prim.* **2**, 51 (2022).
42. Shapson-Coe, A. *et al.* A connectomic study of a petascale fragment of human cerebral cortex. *bioRxiv* 2021.05.29.446289 (2021) doi:10.1101/2021.05.29.446289.
43. Kormacheva, M. *et al.* BIB-mSTEM Approach for Large Scale Acquisition of Brain Tissue. *Microsc. Microanal.* **30**, ozae044.328 (2024).
44. Zheng, Z. *et al.* Fast imaging of millimeter-scale areas with beam deflection transmission electron microscopy. *Nat. Commun.* **15**, 6860 (2024).
45. Ohno, N., Katoh, M., Saitoh, Y., Saitoh, S. & Ohno, S. Three-dimensional volume imaging with electron microscopy toward connectome. *Microscopy* **64**, 17–26 (2015).
46. Song, K., Feng, Z. & Helmstaedter, M. High-contrast en bloc staining of mouse whole-brain and human brain samples for EM-based connectomics. *Nat. Methods* **20**, 836–840 (2023).
47. Hayworth, K. J., Kasthuri, N., Schalek, R. & Lichtman, J. W. Automating the Collection of Ultrathin Serial Sections for Large Volume TEM Reconstructions. *Microsc. Microanal.* **12**, 86–87 (2006).
48. Briggman, K. L. & Bock, D. D. Volume electron microscopy for neuronal circuit reconstruction. *Curr. Opin. Neurobiol.* **22**, 154–161 (2012).
49. Stelzer, E. H. K. *et al.* Light sheet fluorescence microscopy. *Nat. Rev. Methods Prim.* **1**, 73 (2021).
50. Truckenbrodt, S. Expansion Microscopy: Super-Resolution Imaging with Hydrogels. *Anal. Chem.* **95**, 3–32 (2023).
51. Wen, G., Leen, V., Rohand, T., Sauer, M. & Hofkens, J. Current Progress in Expansion Microscopy: Chemical Strategies and Applications. *Chem. Rev.* **123**, 3299–3323 (2023).
52. Cho, Y. *et al.* FRACTAL: Signal amplification of immunofluorescence via cyclic staining of target molecules. *Nanoscale* **12**, 23506–23513 (2020).
53. Saka, S. K. *et al.* Immuno-SABER enables highly multiplexed and amplified protein imaging in tissues. *Nat. Biotechnol.* **37**, 1080–1090 (2019).
54. Wassie, A. T., Zhao, Y. & Boyden, E. S. Expansion microscopy: principles and uses in biological research. *Nat. Methods* **16**, 33–41 (2019).
55. Rathi, P. *et al.* Plasmon-Enhanced Expansion Microscopy. *Nano Lett.* **23**, 5654–5662 (2023).
56. Artur, C. G. *et al.* Plasmonic nanoparticle-based expansion microscopy with surface-enhanced Raman and dark-field spectroscopic imaging. *Biomed. Opt. Express* **9**, 603–615 (2018).
57. Chen, B. *et al.* Lattice light-sheet microscopy: Imaging molecules to embryos at high spatiotemporal resolution. *Science (80-. ).* **346**, (2014).
58. Gao, R. *et al.* Cortical column and whole-brain imaging with molecular contrast and nanoscale resolution. *Science (80-. ).* **363**, eaau8302 (2019).
59. Liu, T.-L. *et al.* Observing the cell in its native state: Imaging subcellular dynamics in multicellular organisms. *Science (80-. ).* **360**, (2018).
60. Tsai, Y.-C. *et al.* Rapid high resolution 3D imaging of expanded biological specimens with lattice light sheet microscopy. *Methods* **174**, 11–19 (2020).
61. Shi, Y., Daugird, T. A. & Legant, W. R. A quantitative analysis of various patterns applied in lattice light sheet microscopy. *Nat. Commun.* **13**, 4607 (2022).
62. Wang, W. *et al.* Nanoscale volumetric fluorescence imaging via photochemical sectioning. *bioRxiv* 2024.08.01.605857 (2024) doi:10.1101/2024.08.01.605857.
63. Park, H.-E. *et al.* Scalable and Isotropic Expansion of Tissues with Simply Tunable Expansion Ratio. *Adv. Sci.* **6**, 1901673 (2019).
64. Chakraborty, T. *et al.* Light-sheet microscopy of cleared tissues with isotropic, subcellular resolution. *Nat. Methods* **16**, 1109–1113 (2019).
65. Prince, M. N. H. *et al.* Signal improved ultra-fast light-sheet microscope for large tissue imaging. *Commun. Eng.* **3**, 59 (2024).
66. Glaser, A. *et al.* Expansion-assisted selective plane illumination microscopy for nanoscale imaging of centimeter-scale tissues. *bioRxiv* 2023.06.08.544277 (2023) doi:10.1101/2023.06.08.544277.

67. Collins, L. T. The case for emulating insect brains using anatomical "wiring diagrams" equipped with biophysical models of neuronal activity. *Biol. Cybern.* **113**, 465–474 (2019).
68. Kuan, A. T. *et al.* Dense neuronal reconstruction through X-ray holographic nano-tomography. *Nat. Neurosci.* **23**, 1637–1643 (2020).
69. Stampfl, A. P. J. *et al.* SYNAPSE: An international roadmap to large brain imaging. *Phys. Rep.* **999**, 1–60 (2023).
70. Bosch, C. *et al.* Non-destructive X-ray tomography of brain tissue ultrastructure. *bioRxiv* 2023.11.16.567403 (2024) doi:10.1101/2023.11.16.567403.
71. Truckenbrodt, S., Sommer, C., Rizzoli, S. O. & Danzl, J. G. A practical guide to optimization in X10 expansion microscopy. *Nat. Protoc.* **14**, 832–863 (2019).
72. Cai, D., Cohen, K. B., Luo, T., Lichtman, J. W. & Sanes, J. R. Improved tools for the Brainbow toolbox. *Nat. Methods* **10**, 540–547 (2013).

**Supplemental information:**
**Comparative prospects of imaging methods for whole-brain mammalian connectomics**

Logan Thrasher Collins,[1,*] Todd Huffman, Randal Koene,[2]

[1]Washington University in St. Louis, Department of Biomedical Engineering
[2]Carboncopies Foundation
[*]Corresponding author

## Supplemental calculations:

## Electron microscopy:

## Yin et al. (2020)[1]

**Raw numbers listed in the paper:**
Total acquisition time: ~6 months
Total volume: ~1 mm$^3$
Number of microscopes running in parallel: 6
Voxel size: 4×40×40 nm
Cost per microscope: $250,000 ($125,000 for microscope plus $125,000 for modifications)

**Imaging speed per microscope (mm$^3$/month) calculation:**
$1\ \mathrm{mm}^3\ /\ 6\ \mathrm{months}\ /\ 6\ \mathrm{microscopes}\ =\ 0.0278\ \mathrm{mm}^3/\mathrm{month}$

**Time to image mouse brain per 1 microscope calculation:**
$500\ \mathrm{mm}^3\ /\ (0.0278\ \mathrm{mm}^3/\mathrm{month})\ =\ 18{,}000\ \mathrm{months}$
$18{,}000/12\ =\ 1{,}500\ \mathrm{years}$

**Time to image human brain per 1 microscope calculation:**
$1{,}200{,}000\ \mathrm{mm}^3\ /\ (0.0278\ \mathrm{mm}^3/\mathrm{month})\ =\ 4.3165 \times 10^7\ \mathrm{months}$
$(4.3165 \times 10^7)/12\ =\ 3.5971 \times 10^6$ years (rounded to 3.6M years)

**Time to image mouse brain with $100M of microscopes calculation:**
Number of microscopes is $\$10^8/\$250{,}000\ =\ 400$ microscopes
$500\ \mathrm{mm}^3/(400 \bullet 0.0278\ \mathrm{mm}^3/\mathrm{month}) = 44.964\ \mathrm{months}$
$(44.964)/12 = 3.747$ years (rounded to 3.75 years)

**Time to image human brain with $100M of microscopes calculation:**
Number of microscopes is $\$10^8/\$250{,}000\ =\ 400$ microscopes
$1{,}200{,}000\ \mathrm{mm}^3/(400 \bullet 0.0278\ \mathrm{mm}^3/\mathrm{month}) = 1.0791 \times 10^5\ \mathrm{months}$
$(1.0791 \times 10^5)/12 = 8{,}992.8$ years (rounded to 9,000 years)

## Shapson-Coe et al. (2021)[2]

**Raw numbers listed in the paper (and elsewhere as noted):**
Total acquisition time: 326 days

Total volume: ~1 $mm^3$
Number of microscopes running in parallel:
Voxel size: 4×40×33 nm
Cost per microscope: $4,000,000, see table 1 of "High-throughput transmission electron microscopy with automated serial sectioning"[3] for multibeam SEM prices.

**Imaging speed per microscope ($mm^3$/month) calculation:**
326 days / 31 = 10.5161 months
$1\ mm^3 / 10.5161\ months / 1\ microscopes = 0.0951\ mm^3/month$

**Time to image mouse brain per 1 microscope calculation:**
$500\ mm^3 / (0.0951\ mm^3/month) = 5{,}257.6\ months$
$5{,}257.6/12 = 438.1353$ years (rounded to 438 years)

**Time to image human brain per 1 microscope calculation:**
$1{,}200{,}000\ mm^3 / (0.0951\ mm^3/month) = 1.2618 \times 10^7$ months
$(1.2618 \times 10^7)/12 = 1.0515 \times 10^6$ years (rounded to $1.05 \times 10^6$ years)

**Time to image mouse brain with $100M of microscopes calculation:**
Number of microscopes is $\$10^8/\$4{,}000{,}000 = 25$ microscopes
$500\ mm^3/(25 \bullet 0.0951\ mm^3/month) = 210.3049$ months
$(210.3049)/12 = 17.5254$ years (rounded to 17.5 years)

**Time to image human brain with $100M of microscopes calculation:**
Number of microscopes is $\$10^8/\$4{,}000{,}000 = 25$ microscopes
$1{,}200{,}000\ mm^3/(25 \bullet 0.0951\ mm^3/month) = 5.0473 \times 10^5$ months
$(5.0473 \times 10^5)/12 = 42{,}061$ years

## Zheng et al. (2021)[4]

**Raw numbers listed in the paper (and elsewhere as noted):**
Projected acquisition time to image 1 $mm^3$ with 4 microscopes in parallel: 37 days (assuming 65% uptime)
Number of microscopes running in parallel: 4
Voxel size: 3.6×3.6×45 nm
Cost per microscope: $500,000

**Imaging speed per microscope ($mm^3$/month) calculation:**
37 days / 31 = 1.1935 months for 1 $mm^3$
$1.1935\ months/mm^3 / 4\ microscopes = 0.2984\ months/mm^3/microscope$
(rounded to 0.3 $mm^3$/month for 1 microscope)

**Time to image mouse brain per 1 microscope calculation:**
$500\ mm^3 / (0.3\ mm^3/month) = 1{,}666.7$ months
$1{,}666.7/12 = 139.6336$ years (rounded to 139.6 years)

**Time to image human brain per 1 microscope calculation:**
$1{,}200{,}000\ mm^3 / (0.3\ mm^3/month) = 4{,}000{,}000$ months
$4{,}000{,}000/12 = 333{,}333.33$ years (rounded to 333,333 years)

**Time to image mouse brain with $100M of microscopes calculation:**
Number of microscopes is $\$10^8/\$500{,}000 = 200$ microscopes
$500\ mm^3/(200 \bullet 0.3\ mm^3/month) = 8.333$ months
$(8.333)/12 = 0.6944$ years (rounded to 0.69 years)

**Time to image human brain with $100M of microscopes calculation:**
Number of microscopes is $\$10^8/\$500{,}000 = 200$ microscopes
$1{,}200{,}000\ mm^3/(200 \bullet 0.3\ mm^3/month) = 20{,}000$ months
$(20{,}000)/12 = 1{,}666.7$ years (rounded to 1,667 years)

## Expansion light-sheet fluorescence microscopy

## Wang et al. (2024)

**Raw numbers listed in the paper (and elsewhere as noted):**
Number of microscopes running in parallel: 1
Voxel size: 56×56×135 nm
Cost per microscope: $300,000 as found through personal correspondence with the authors.
Acquisition speed: 9.4×6.4×4.3 mm (after 2-fold expansion) per 10 days

**Effective voxel size with 7-fold expansion[5] calculation:**
$56 \times 56 \times 135\ nm\ /4.5 = 12.44 \times 12.44 \times 30$ nm voxels $= 4{,}645.6\ nm^3$ volume per voxel

**Imaging speed per microscope ($mm^3$/month) calculation**
$(9.4)(6.4)(4.3)\ mm^3/(10\ days) = 1 \bullet (25.87\ mm^3/day)(31$ days per month)
$= 801.97\ mm^3/$ month (rounded to $802\ mm^3/month$)

**Time to image 4.5-fold expanded mouse brain per 1 microscope calculation:**
$(500\ mm^3 \bullet 4.5^3)\ /\ (801.97\ mm^3/month) = 56.81$ months
$56.81/12 = 4.734$ years (rounded to 4.7 years)

**Time to image 4.5-fold expanded human brain per 1 microscope calculation:**
$(1{,}200{,}000\ mm^3 \bullet 4.5^3)\ /\ (801.97\ mm^3/month) = 136{,}350$ months
$(136{,}350)/12 = 11{,}363$ years

**Time to image 4.5-fold expanded mouse brain with $100M of microscopes calculation:**
Number of microscopes is $\$10^8/\$300{,}000 = 333$ microscopes
$(500\ mm^3 \bullet 4.5^3)\ /\ (333$ microscopes $\bullet\ 801.97\ mm^3/month) = 0.171$ months
$0.171/12 = 0.0142$ years (rounded to 0.014 years)

**Time to image 4.5-fold expanded human brain with $100M of microscopes calculation:**
Number of microscopes is $\$10^8/\$300{,}000 = 333$ microscopes
$(1{,}200{,}000\ mm^3 \bullet 4.5^3)\ /\ (333$ microscopes $\bullet\ 801.97\ mm^3/month) = 409.46$ months
$409.46/12 = 34.12$ years (rounded to 34 years)

## Chakraborty et al. (2019)[6]

**Raw numbers listed in the paper (and elsewhere as noted):**
Number of microscopes running in parallel: 1

Voxel size: 425×425×425 nm, see passage in Chakraborty et al.'s Methods "*an isotropic step size of 0.425 µm and 18% overlap between stacks for faithful stitching*" and their Supplementary Table 6.
Cost per microscope: $120,000; see Chakraborty et al.'s Supplementary Table 5.
Acquisition speed: 17.6 minutes per $mm^3$, see passage in Chakraborty et al.'s Methods "*Imaging a 1-$mm^3$ volume using ctASLM took 17.6 min*".

**Effective voxel size with 24-fold expansion[7] calculation:**
$425 \text{ nm isotropic}/24 = 17.7 \text{ nm isotropic voxels}$

**Imaging speed per microscope ($mm^3$/month) calculation:**
$1 \text{ mm}^3/17.6 \text{ minutes} = 1 \bullet (44{,}640 \text{ minutes per month } / \ 17.6)$
$= 2{,}536.36 \text{ (rounded to } 2{,}536 \text{ mm}^3\text{/month)}$

**Time to image 24-fold expanded[7] mouse brain per 1 microscope calculation:**
$(500 \text{ mm}^3 \bullet 24^3) \ / \ (2{,}536 \text{ mm}^3\text{/month}) \ = \ 2{,}725.6 \text{ months}$
$2{,}725.6/12 \ = \ 227.1293 \text{ years (rounded to 227 years)}$

**Time to image 24-fold expanded[7] human brain per 1 microscope calculation:**
$(1{,}200{,}000 \text{ mm}^3 \bullet 24^3) \ / \ (2{,}536 \text{ mm}^3\text{/month}) \ = \ 6.5413 \times 10^6 \text{ months}$
$(6.5413 \times 10^6)/12 \ = \ 5.4511 \times 10^5 \text{ years (rounded to } 5.45 \times 10^5 \text{ years)}$

**Time to image 24-fold expanded[7] mouse brain with $100M of microscopes calculation:**
$\text{Number of microscopes is } \$10^8/\$120{,}000 \ = \ 833 \text{ microscopes}$
$(500 \text{ mm}^3 \bullet 24^3) \ / \ (833 \text{ microscopes } \bullet \ 2{,}536 \text{ mm}^3\text{/month}) \ = \ 3.272 \text{ months}$
$3.272/12 \ = \ 0.2727 \text{ years (rounded to 0.273 years)}$

**Time to image 24-fold expanded[7] human brain with $100M of microscopes calculation:**
$\text{Number of microscopes is } \$10^8/\$120{,}000 \ = \ 833 \text{ microscopes}$
$(1{,}200{,}000 \text{ mm}^3 \bullet 24^3) \ / \ (833 \text{ microscopes } \bullet \ 2{,}536 \text{ mm}^3\text{/month}) \ = \ 7{,}852.7 \text{ months}$
$7{,}852.7/12 \ = \ 654.3942 \text{ years (rounded to 654 years)}$

## Prince et al. (2023)[8]

**Raw numbers listed in the paper (and elsewhere as noted):**
Number of microscopes running in parallel: 1
Voxel size: 425×425×425 nm, see passage in Prince et al.'s paper "*The axial step size is determined by the lateral pixel size at the image space. For instance, the measured magnification of SIFT in water is 15.28x, which gives a lateral image pixel size of 0.425 µm*" and their Supplementary Table 3. Slightly better magnifications were possible in other immersion media, but we have chosen to use the value for water since expanded tissues are typically filled with pure water.
Cost per microscope: $104,000 as found through personal correspondence with the authors.
Acquisition speed: 5.5×3.6×3.5 $mm^3$ / 13.91 hours, see passage in Prince et al.'s paper "*although the total pfAT of imaging 5.5 × 4.6 × 3.5 $mm^3$ mouse hind-paw for SIFT and traditional ASLM are 7.63 and 30.55 hours respectively, the total imaging time for SIFT and traditional ASLM are 13.91 and 40.48 hours respectively*".

**Effective voxel size with 24-fold expansion[7] calculation:**
$425 \text{ nm isotropic}/24 = 17.7 \text{ nm isotropic voxels}$

**Imaging speed per microscope ($mm^3$/month) calculation:**

$(5.5)(3.6)(3.5)\ \text{mm}^3/13.91\ \text{hours} = 69.3\ \text{mm}^3/\ 13.91(60) = 0.083\ \text{mm}^3/\text{minute}$
$= (0.083\ \text{mm}^3)(44{,}640\ \text{minutes per month}) = 3{,}706.6\ (\text{rounded to } 3{,}707\ \text{mm}^3/\text{month})$

**Time to image 24-fold expanded[7] mouse brain per 1 microscope calculation:**
$(500\ \text{mm}^3 \bullet 24^3)\ /\ (3{,}707\ \text{mm}^3/\text{month}) = 1{,}864.6\ \text{months}$
$1{,}864.6/12 = 155.3817\ \text{years (rounded to 155 years)}$

**Time to image 24-fold expanded[7] human brain per 1 microscope calculation:**
$(1{,}200{,}000\ \text{mm}^3 \bullet 24^3)\ /\ (3{,}707\ \text{mm}^3/\text{month}) = 4.475 \times 10^6\ \text{months}$
$(4.475 \times 10^6)/12 = 3.7292 \times 10^5\ \text{years (rounded to } 3.73 \times 10^5\ \text{years)}$

**Time to image 24-fold expanded[7] mouse brain with $100M of microscopes calculation:**
Number of microscopes is $\$10^8/\$104{,}000 = 962$ microscopes
$(500\ \text{mm}^3 \bullet 24^3)\ /\ (962\ \text{microscopes} \bullet 3{,}707\ \text{mm}^3/\text{month}) = 1.9382\ \text{months}$
$1.9382/12 = 0.1615\ \text{years (rounded to 0.162 years)}$

**Time to image 24-fold expanded[7] human brain with $100M of microscopes calculation:**
Number of microscopes is $\$10^8/\$120{,}000 = 962$ microscopes
$(1{,}200{,}000\ \text{mm}^3 \bullet 24^3)\ /\ (962\ \text{microscopes} \bullet 3{,}707\ \text{mm}^3/\text{month}) = 4{,}651.8\ \text{months}$
$4{,}651.8/12 = 387.6467\ \text{years (rounded to 388 years)}$

**Supplemental references:**

1. Yin, W. *et al.* A petascale automated imaging pipeline for mapping neuronal circuits with high-throughput transmission electron microscopy. *Nat. Commun.* **11**, 4949 (2020).
2. Shapson-Coe, A. *et al.* A connectomic study of a petascale fragment of human cerebral cortex. *bioRxiv* 2021.05.29.446289 (2021) doi:10.1101/2021.05.29.446289.
3. Graham, B. J. *et al.* High-throughput transmission electron microscopy with automated serial sectioning. *bioRxiv* 657346 (2019) doi:10.1101/657346.
4. Zheng, Z. *et al.* Fast imaging of millimeter-scale areas with beam deflection transmission electron microscopy. *bioRxiv* 2022.11.23.517701 (2022) doi:10.1101/2022.11.23.517701.
5. Park, H.-E. *et al.* Scalable and Isotropic Expansion of Tissues with Simply Tunable Expansion Ratio. *Adv. Sci.* **6**, 1901673 (2019).
6. Chakraborty, T. *et al.* Light-sheet microscopy of cleared tissues with isotropic, subcellular resolution. *Nat. Methods* **16**, 1109–1113 (2019).
7. M'Saad, O. *et al.* All-optical visualization of specific molecules in the ultrastructural context of brain tissue. *bioRxiv* 2022.04.04.486901 (2022) doi:10.1101/2022.04.04.486901.
8. Prince, M. N. H. *et al.* Signal Improved ultra-Fast Light-sheet Microscope (SIFT) for large tissue imaging. *bioRxiv* 2023.05.31.543002 (2023) doi:10.1101/2023.05.31.543002.